\documentstyle[12pt]{article}
\def\TwoColumn{\par\global\columnwidth\textwidth
   \global\advance\columnwidth -\columnsep \global\divide\columnwidth\tw@
   \global\hsize\columnwidth \global\linewidth\columnwidth
   \global\@twocolumntrue \global\@firstcolumntrue
   \@dblfloatplacement\@ifnextchar[{\@topnewpage}{}}
\def\eps{\epsilon }
\def\epsbr{\overline{\epsilon} }

\def\Ibf{{\bf I}\,}

\def\sbf{{\bf{s}}\, }

\def\mubf{{\bf{\mu}}\, }

\def\hh{\hat{h} }
\def\ebr{\overline{e} }

\def\xb{\overline{x} }

\def\phl{{ \phi_{\ell}}\, }
\def\phlp{{ \phi_{\ell'}}\, }
\def\phtl{\tilde{ \phi}\, }
\def\mubf{{\bf \mu}\, }

\def\om{{ \omega}\, }
\def\oml{{ \omega_{\ell}}\, }
\def\omo{{ \omega_{o}}\, }
\def\omlp{{ \omega_{\ell'}}\, }
\def\Sl{{\sigma^2}\, }

\def\sgb{\overline{\sigma}}
\def\pb{\overline{p}}

\def\Rbf{{\bf R}\,}

\def\Rbr{{\bf \bar{R}}\,}
\def\muR{{\bf \mu}\, }
\def\muRb{{\bf \mu}\, }

\def\hb{\overline{h} }

\def\sgph{\sigma_{\phi}^2}

\def\ns{\vskip 2pc\noindent}       

\def\IR{I\kern-.255em R}

\def\eps{\epsilon }

\def\Sbf{{\bf S}\, }

\def\Sbf{{\bf S}\, }
\def\albf{{\bf{\alpha}}\, }

\def\mubf{{\bf{\mu}}\, }

\def\eb{{\bf e\,}}
\def\xb{{\bf x\,}}
\def\zb{{\bf z\,}}

\def\eps{ \epsilon}

\def\ebf{{\bf e}\,}

\def\Ibf{{\bf I}\,}

\def\Sbf{{\bf S}\,}

\def\ytl{\tilde{y}}
\def\ztl{\tilde{z}}
\def\siml{\underline{\sim}}
\def\cl{\centerline}

\def\bm{\boldmath}
\def\part{\partial}

\def\Abr{\bar{A}}

\def\Sbf{{\bf{S}}}

\def\cl{\centerline}

\def\Sbf{{\bf{S}}}

\begin{document}
\begin{center}
{\bf Kernel Estimation of the Instantaneous Frequency}
\end{center}
\begin{center}
{Kurt S. Riedel \\
Courant Institute of Mathematical Sciences \\
New York University \\
New York, New York 10012-1185 }\\
\end{center}

\begin{abstract}
We consider kernel estimators of the instantaneous frequency of
a slowly evolving sinusoid in white noise. The expected estimation error
consists of two terms. The systematic bias error grows as the kernel 
halfwidth increases while the random error decreases. 
For a nonmodulated signal, $g(t)$, the kernel halfwidth which 
minimizes the expected error scales as
$h \sim \left[{ \sigma^2 \over 
 N| \partial_t^2 g^{}|^2 } \right]^{1/ 5}$,
where 
$\sigma^2$ is the noise variance and $N$ is the number of measurements
per unit time.
We show that estimating the instantaneous frequency corresponds to
estimating the  first derivative  of a modulated signal, $A(t)\exp(i\phi(t))$.
For instantaneous frequency estimation,
the halfwidth which minimizes the expected  error is larger:
$h_{1,3} \sim  \left[{ \sigma^2 \over 
A^2N| \partial_t^3 (e^{i\phtl(t)} )|^2 } \right]^{1/ 7}$.
Since the optimal halfwidths depend on derivatives of the unknown function,
we initially estimate these derivatives prior to estimating the actual signal.
\end{abstract}
\newpage
\noindent
\cl{I. INTRODUCTION}

We consider the problem of estimating the instantaneous frequency of
one or more slowly evolving sinusoids in white noise. 
Excellent reviews of the estimation of the instantaneous frequency
as well as the general theory of time--frequency distributions 
can be found in [2,3,10]
Cohen \& Lee [4] determine an optimal kernel smoother by minimizing the
time--frequency spread of the resulting estimate of the 
instantaneous frequency. 

Our approach is based on the theory of kernel smoothers for nonparametric
function and derivative estimation [7-9,13-15,17,18,21]. 
Kernel smoothers are weighted averages of the measured values of a slowly
evolving unknown function. We use ``kernel smoother'' to be consistent with
the terminology of nonparametric function estimation. In the electrical
engineering literature, the equivalent terminology is ``linear transfer
function'' or ``acausal finite impulse response linear filter.''
As the kernel 
halfwidth increases, the random error from the white noise decreases.

Lovell \& Williamson (L \& W) [12]
use a centered difference to estimate the time derivative of the phase,
$\phi'(t)$, and then
use a kernel smoother to reduce the variance of the estimate of $\phi'(t)$.
As in L \& W, we treat estimation of  the instantaneous frequency 
as a kernel smoothing problem with circular statistics.  
However, the kernel smoother has a bias error from systematic evolution
of the amplitude and frequency, and this bias error increases with the 
kernel halfwidth.  
We calculate the leading order expected estimation error  by
expanding the unknown function in the ratio of the sampling time to
the characteristic time scale on which the unknown signal is evolving. 
We then  determine the optimal kernel
halfwidth by minimizing the expected error.
In L \& W's pioneering work, the bias error is neglected, and as a result
their estimate of the expected error is a monotonically decreasing function 
of the kernel halfwidth.

A second group of estimators of the instantaneous frequency have 
been developed which are based on linear regression 
or linear prediction [6,11,24]. 
These methods estimate fixed frequencies. Since the frequencies are
assumed to be time independent,  and these methods neglect
the rate of change of the frequencies, and the temporal  evolution of the
signal frequencies causes a bias error in the estimate. Often, these
methods use short subsequences such that on a particular subsequence
the bias error (from frequency evolution) is negligible.
However, having negligible bias is really a disadvantage 
because the subsequence length could be increased until the bias error
is comparable with the random error. 
In our approach,
we try to minimize the total error by increasing the kernel halfwidth 
until the rate of increase in the bias error matches the decrease in the
variance.

In the next section, we review the theory of nonparametric
function and derivative estimation. In Section III, we apply these
results to instantaneous frequency estimation. In Section IV,
we generalize the analysis to include the correlated errors which are
induced by the Hilbert transform. In Section V, we consider multiple signals.
In  Appendix A,
we describe data-adaptive multiple stage kernel estimators which determine
a self-consistent optimal halfwidth. In Appendix B, we describe
the kernel shapes which minimize the expected error. 

\ \\

\cl{II. EXPECTED LOSS OF KERNEL SMOOTHERS}
\vspace{.1in}

In this section and the appendix, we consider
a real digital signal 
in white noise:
$$y_j = g(t_j)  + \tilde{e}_j
\ \ , \ \  j=1,\ldots N , \eqno (2.1)$$
where $\tilde{e}_j$ is independently distributed noise 
with variance $\sigma^2$.
Our goal is to estimate the $q$th derivative of $g(t_j)$ with 
a minimum of expected error. 
We assume that $g(t)$ has  $\pb$ continuous derivatives
and that $g(t)$ varies slowly with respect to the sampling rate.
We normalize the measurement times, $t_j$, to be in
the closed interval $[0,1]$.

We consider kernel estimators of $\partial_t^{q}g(t)$ of the form:
$$
\widehat{\partial_t^{q}g}(t)  = {1\over N h^{(q+1)}}\sum_{j=1}^{N}  
K({t-t_j \over h}) {y}_j 
\ ,\eqno(2.2)$$
where the $\widehat{ \ }$ over ${\partial_t^{q}g}$ denotes the 
estimate of the $q$th derivative.
We define the  vector, 
$\mu(t) ={1\over N h}(K({t-t_1\over h})), \ldots K({t-t_N\over h}))^T$.
We say a kernel, $\mubf$, with halfwidth, $h$, is of order $(q,p)$ if
$$
\mubf \cdot\sbf^{(m)} =  { q!\ }\delta_{m,q} \ , \ m= 0, \ldots,p-1
,\eqno(2.3)$$
where
$\sbf^{(m)}_{j} \equiv ({t- t_j\over h})^{m}$. 
We denote the $p$th moment of a kernel of order $(q,p)$ by $C_{q,p}$:
$\mubf \cdot\sbf^{(p)} =   p!\ C_{q,p}$.
Kernels of order $(q,p)$ are used to estimate the $q$th derivative
of the function to order $O(h^{p-q})$. 
We normally select $p= q+2$ and our preferred set of kernels 
is given in (2.10). For function
estimation ($q=0$), we normally use $p=2$ and occasionally use $p=4$.
To estimate the instantaneous frequency, we use a kernel smoother of
order $(1,3)$.

The moment conditions (2.3) are also satisfied by the phase difference
estimators of Boashash [2]. Boashash's estimators are chosen to have
the shortest possible length: $N= p+1$.  As a result, these phase difference
estimators have near minimal bias error and are suitable for high 
resolution estimates in a noiseless signal. If noise is present, these
phase difference estimators will appreciably amplify the noise.
In a noisy signal, our kernel  estimators reduce the variance by 
averaging over many more data points than the kernel order, $p$.

The variance of the kernel estimator is
$$
{\rm Var} \ [ \widehat{g^{(q)}}(t,\mubf)] = {\sigma^2 m_{2}(\mu)
\over  N h^{2q+1}}
\  \ ,\eqno (2.4)$$
where $m_{2}(\mu)=  ||\mubf||^2 \times (Nh) \siml \int K(s)^2 ds$.
Expanding $g(t_j)$ in a  Taylor series  about $g(t)$,
the bias of a kernel smoother of order $(q,p)$  is
$$E\left[ \widehat{\partial_t^{q}g} (t)\right]-  \partial_t^{q}g(t) =  
C_{q,p} \partial_t^{p} g(t) h^{p-q} 
.\eqno(2.5)$$
The leading order total squared error of $\partial_t^q g(t_j)$ is
$$L^2(\widehat{\partial_t^q g(t_j)};\mubf ) =
C_{q,p}^2 |\partial_t^p g(t_j)|^2  h^{2(p-q)}
+ {\Sl m_{2}(\mu) \over N h^{2q+1}} \
,\eqno(2.6)$$
where 
the corrections are ${\cal O}(h^{2(p-q)+1})$.
Solving (2.6) for the optimal value
of the kernel scale size yields
$$h_{o}(\mubf ) = \left[ {2q+1 \over 2(p-q)}
{ \Sl m_{2}(\mu) \over  C_{q,p}^2 N |\partial_t^p g(t_j)|^2 }
\right]^{1\over 2p+1}
\ .\eqno(2.7)$$
For this choice of kernel width, $h_o$, the total squared error of (2.2) is
proportional to
$$
L^2(\widehat{\partial_t^q g}(t_j) ) \ \siml \
M_{q,p}  |C_{q,p}\partial_t^p g(t_j)|^{2(2q+1)\over(2p+1)}
\left({ \Sl m_{2}(\mu) \over N } \right)^{2(p-q)\over(2p+1)}
\ , \eqno(2.8)$$
where $M_{q,p}\equiv ( {2q+1 \over 2(p-q)} )^{2(p-q)\over(2p+1)} +
( {2(p-q) \over (2q+1)} )^{(2q+1)\over(2p+1)}$.
The optimal $h$ is proportional to $N^{-1/(2p+1)}$,
and the total squared error, $ L^2(\partial_t^q g)$,
is proportional to $N^{-2(p-q)\over(2p+1)}$.
If  $g(t)$ has $\overline{p}$  continuous derivatives, where
$q \le \overline{p} \le p$, the optimal bandwidth scales as  
$N^{-1/(2\overline{p}+1)}$, and the total squared error
is proportional to $N^{-2(\overline{p}-q)/(2\overline{p}+1)}$.
This convergence rate 
is optimal for functions with precisely $\overline{p}$ continuous derivatives
[22]. 

In [21] and Appendix B, we evaluate the kernel shape 
(under appropriate constraints) which minimizes the expected error.
In the high sampling rate limit, the kernel shapes which minimize
the local expected loss (as given by (2.6)) are
independent of the kernel halfwidth and can be explicitly evaluated.
(See bibliography in [21].) For $p= q+2$, the limiting shape of
the optimal kernel is 
$$K(t) = \gamma \left[P_q(t)-P_{q+2}(t)\right] \ ,
\eqno (2.9)$$
where $P_q(t)$ and $P_{q+2}(t)$ are the Legendre polynomials 
(or their discrete analog)
and $\gamma  \equiv \prod_{k=1}^{q+1}\frac{(q+k)}{2}$.
In (2.9), the
estimation point is at $t = 0$ and the kernel support is $[-1,1]$.
For  $q=0$, Eq.~(2.9) reduces to $K(t)= \frac{3}{4}(1-t^2)$,
and for  $q=1$, $K(t)= \frac{15}{4}(t-t^3)$.
When the domain of the kernel smoother intersects the ends of the
dataset, the kernel requires the more general form: $K(t,{t-t_j \over h})$
to continue to be of order $(q,p)$. The appropriate edge kernels are given in 
[21]. 

We have derived the optimal kernel 
halfwidth assuming that  $g^{(p)}(t)$ is known. 
In practice, $g^{(p)}(t)$ is unknown and needs to be estimated.
In the appendix, we describe data adaptive methods where 
we estimate $\part_t^p g(t)$ using a higher order kernel of
order $(p,p+2)$ and substitute $\widehat{\part_t^p g}(t)$ into (2.7).
More detailed treatments of kernel estimation
can be found in [7-9,13-15,17,18,21].  

\ \\

\cl{III. INSTANTANEOUS FREQUENCY ESTIMATES}
\vspace{.05in}

We now consider models where both the amplitude, $A(t)$, and
the instantaneous frequency, $\phi'(t)$, are evolving  slowly with
respect to both the sampling rate and the characteristic oscillation
frequency, $\omega_o$. The measured data satisfies
$$y_j = A(t_j) \cos( \phi(t_j)) + \tilde{e}_j
\ , \ \  j=1,\ldots N , \eqno (3.1)$$
where $\tilde{e}_j$ is i.i.d. noise with variance $\sigma^2$.
We define the analytic signal, $\zb = A[\xb] =\xb + i {\rm H}[\xb]$,
where H is the Hilbert transform.
We assume that $\phi(t)= \omega_o t + \phi_o + \phtl(t)$,
where the characteristic frequency, $\omega_o$, is given. 
In practice, we iterate on $\omega_o$
with  the new value of $\omega_o$ being the previous estimate of the 
instantaneous frequency.

We assume that $A(t)$ and $\phi(t)$
satisfy the bandwidth conditions [2]: 
${\cal F}[A(t)]$ vanishes for $|\omega| > \omega_b $ 
and ${\cal F}[cos(\phi(t))]$ vanishes for $|\omega| < \omega_b $, 
where ${\cal F}$ is the Fourier transform 
and $\omega_b$ is fixed.
In this case, the analytic signal has the phasor representation:
$$z_j = A(t_j) \exp( i \phi(t_j)) + {\eps}_j
\ , \ \  j=1,\ldots N , \eqno (3.2)$$
where ${\bm \eps} \equiv A[\eb]$. The Hilbert transform couples the
$\eps_j$ so that they are not independent. We initally consider
the case when the model of (3.2) holds with 
${\rm Re}[\eps_j]$ and ${\rm Im}[\eps_j]$ as independent
random variables with variance $\sgb^2$.
In this case, the exact distribution of the phase is known. 
Because the distribution of
$\exp(i \arg(z_j))$ is more nearly Gaussian
distributed than is $\arg(z_j)$ [23], 
we smooth and differentiate $\exp(i \arg(z_j))$
instead of $\arg(z_j)$.

To apply the optimal kernel smoother theory of Sec.~II to (3.2),
we require that the sampling rate is fast with respect to the
characteristic evolution time. To remove the $\exp(i\omega_ot)$
modulation from the sampling rate constraint, we demodulate
the data about the central frequency. We define
$\ztl^{}_j \sim z_j e^{-i(\omega_o t+\phi_{o})}$.
By kernel smoothing $\ztl_j$, we can  estimate the
real and imaginary parts of $A(t)e^{i\phtl_{}(t)}$.
However, for time-frequency representations, we need
the modulus, $A$, and instantaneous frequency, ${\phi_{}'(t)}$.
We now describe our kernel estimation scheme for the 
instantaneous frequency.

We begin by computing $u_j = \exp(i \arg(\ztl_j))$. Provided that the
signal to noise ratio, $s ={ A^2 \over 2\sgb^2}$, is high, $u_j$ is 
approximately distributed as $u_j \sim {\cal N}[\exp(i \phtl(t_j),\sgph]$,
where $\sgph = \frac{\sgb^2}{A^2+\sgb^2}$.
We then estimate $e^{i\phtl_{}(t)}$ and $\partial_t e^{i\phtl_{}(t)}$ 
using kernel smoothers of orders (0,2) and (1,3) respectively.
To apply (2.7) \& (2.8), we make the substitutions:
$e^{i\phtl_{}(t)} \rightarrow g(t)$ and 
$\sgph = \frac{\sgb^2}{A^2+\sgb^2} \rightarrow \sigma^2$.
 
Using these results, our instantaneous frequency estimate is
$$\widehat{\phi_{}'}(t) = \omega_o + Im\left[
{\widehat{\part_t e^{i\phtl}}~^{(t)}  \over 
\widehat{e^{i\phtl}}~^{(t)}  }
\right] ,\eqno (3.3)$$
Since our initial guess of the centering frequency, $\omega_o$,
may be inaccurate, we can iterate the local kernel estimates by replacing
$\omega_o$ with $\omega_o + \widehat{\phtl_{}'}(t)$. 
The error estimates for $\widehat{\phi_{}'}$ are dominated
by the error in  $\widehat{\part_t e^{i\phtl}}~^{(t)}$. Thus
the optimal halfwidth to estimate $\part_t e^{i\phtl(t)}$ is
$$h_{o}(\mubf ) = \left( {3 \over 4}
{ \sgph m_{2}(\mu) \over  C_{1,3}^2 N |\partial_t^3 e^{i\phtl}|^2 }
\right)^{1\over 7} \ \ . \eqno(3.4)$$ 
The resulting error in $\widehat{\partial_t\phi_{}}(t)$ is
$$
L^2(\widehat{\partial_t \phi}(t_j) ) \ \siml \
M_{1,3}  |C_{1,3}\partial_t^3 e^{i\phtl} |^{6 \over 7}
\left({ \sgph m_{2}(\mu) \over N } \right)^{4 \over 7}
\ . \eqno(3.5)$$

Our work generalizes the kernel estimators of Lovell \& Williamson [12]
by including bias in estimate of the error and by applying the resulting 
optimal kernel theory. 
L \& W extract pointwise estimates of $\phi'(t_j)$ from
$\arg(z_{j+1}) -\arg(z_{j-1})$ and then smooth these estimates.
We have partially reversed the order of the smoothing and nonlinear
transformations by working with $\exp(i \arg(z_j))$. 
Our basic algorithm is to choose the kernel smoother/derivative estimator
by minimizing the expected loss including the bias error.
This algorithm can also be applied to other orderings of the
smoothing and nonlinear transformations such as that of L \& W. 
However, the distribution of $\exp(i \arg(z_j))$ matches the hypotheses of
kernel smoothing better than most other choices [23]. 

\ \\

\cl{IV. EVOLVING SINUSOIDS IN COLORED NOISE}

\vspace{.05in}

In Section III, we noted that the Hilbert transformed
noise is correlated. To treat this situation, we review kernel estimation
with an arbitrary covariance structure [A]: 
${\rm Cov}[e_j,e_k] = \sigma^2R_{j,k}$. In this case, the variance of the 
kernel estimate of the demodulated data (generalizing (2.4)) is
$$
{\rm Var} \ [ \widehat{g^{(q)}}(t)] = {\sigma^2 \over  N h^{2q+1}}
\sum_{j,k=1}^{N} \mubf_j \Rbf_{j,k} \mubf_k e^{i{\omega_o}(k-j)} 
\  \ .\eqno (4.1)$$
When the errors are autocorrelated, $R_{j,k}= R(j-k)$,
(4.1) can be reformulated in the frequency domain as
$${\rm Var} \ [ \hat{g}(t)] =
{1\over 2\pi h^{2q}} \int_{-\pi}^{\pi}  S(\om)|U(\om-\omega_o)|^2d\om 
\ ,\eqno (4.2)$$
where $U(\omega)$ is the Fourier transform of $\mu$.
If $U(\omega)$ is localized near the zero frequency,
the variance of the estimate will depend almost exclusively 
on the spectral density near $\omo$, $S(\omo)$.
Thus
${\rm Var} \ [ \widehat{g^{(q)}}(t;\mubf)] = S(\omo)  m_2(\mu) / N h^{2q+1}$.

Returning to the Hilbert transform problem, 
we use
the covariance structure of the Hilbert transform:
${\rm Cov}[\eps_j,\epsbr_k] = 
2\sigma^2 \left( \delta_{j,k} + i \xi(j-k)\right)$
and ${\rm Cov}[\eps_j,\eps_k] = 0$, where 
$\xi(j-k) = {2\over \pi (j-k)}$ if $j-k$ is odd and zero otherwise.  
Thus ${\rm E}[|{\rm Re}(\eps_j)|^2] = 
{\rm E}[|{\rm Im}(\eps_j)|^2] = \sigma^2$.
Equation (4.2) shows that ${\rm Im}(\Rbf)$ does not contribute to the variance
of the kernel estimate of  ${A}(t) e^{{i\phtl}(t)}$.
Thus ${\rm Var}[ \widehat{A(t) e^{i\phtl}}] =
2\sigma^2 m_2(\mu) / N h^{2q+1}$. The factor of two arises 
because we are estimating both the real and imaginary parts.

In kernel smoothing $\exp(i \arg(z_j))$, the phase error are distributed
as 
$$e_{\phi,j} \ \siml\
{e^{i(\phi(t_j)+\pi)}{\rm Im}[\eps_j e^{-i\phi(t_j)}]\over
\sqrt{A^2 +\sigma^2}}\ \ .\eqno (4.3)$$
Thus $e_{\phi,j}$ has an approximately normal distribution 
with covariance [12]:
$${\rm Cov}[e_{\phi,j},\ebr_{\phi,k} ] = 
{\sigma^2 \over {A^2 + \sigma^2}} \left[ \delta_{j,k} - \xi(j-k) 
 e^{i(\phi(t_j)-\phi(t_k))}\sin\left(\phi(t_j)-\phi(t_k)\right) 
\right]\ \ .\eqno (4.4)$$
We substitute (4.4) for (2.4)
and apply the resulting kernel halfwidths to (3.3).

\ \\

\cl{V. MULTIPLE EVOLVING SINUSOIDS}
\vspace{.05in}

We now consider signals which consist of a sum of slowly evolving sinusoids in
colored noise:
$$y_j = \sum_{\ell =1}^L A_{\ell}(t_j)  \cos(\phi_{\ell}(t_j))
+ \tilde{e}_j
\ \ . \eqno (5.1)$$
We assume that $\phl(t)= \oml t + \phi_{o,\ell} + \phtl_{\ell}(t)$,
where the frequencies, $\oml$, are given and distinct. 
We require the bandwidth conditions:    
${\cal F}[A_{\ell}(t)]$ vanishes for $|\omega| > \omega_b $ 
and ${\cal F}[cos(\phi_{\ell}(t))]$ vanishes for 
$|\omega-\omega_{\ell} | < \omega_b $, where $\omega_b$ is fixed.
Furthermore, we assume that the supports of
${\cal F}[A_{\ell}(t)cos(\phi_{\ell}(t))], \ell = 1,\ldots L$, 
have negligible overlap.

We initially estimate $\widetilde{A_{\ell}\cos(\phl)}, \ \ell= 1,\ldots L$,
ignoring the bias from  the other sinusoids. 
We then 
attempt to remove the effect of the other sinusoids.
Given estimates, $\widehat{A_{\ell'}\cos(\phl)}(t)$, of the other sinusoids, 
we define the $\ell$th corrected data set, $\ytl^{(\ell)}$, to be
$$
\ytl^{(\ell)}_j \equiv y_j - 
 \sum_{\stackrel{\ell^{\prime} =1}{\ell' \neq \ell} }^L
\widehat{A_{\ell'}\cos\left(\phlp(t_j)\right)}
\ \  . \eqno(5.2)$$
We 
estimate the $\ell$th instantaneous frequency using the corrected dataset.
The correction and  estimation may be iterated. 

To determine when we can neglect the bias from the other line frequencies,
we compare the relative size of the bias from the time evolution of 
$A_{\ell}$, with coherent interference from the other line frequencies.
We denote $\widetilde{A_{\ell'}\cos(\phlp)} \equiv
A_{\ell'}\cos(\phlp)-\widehat{A_{\ell'}\cos(\phlp)}$, 
which is the error in the kernel estimate
of ${A_{\ell'}\cos(\phlp)}$.             
We assume that the  support of the Fourier transform of
$\widetilde{A_{\ell'}\cos(\phlp)}$ 
is contained in the interval $[-\omega_b,\omega_b ]$.
We can neglect the interference of the $\ell'$ line in (2.7) if
$$\left| C_{q,p} \partial_t^{p} A_{\ell} (t) h^p \right| \ >> \
\left| \tilde{A}_{\ell'}(t) U(|\oml - \omlp| - \omega_b ) \right|
\ ,\eqno(5.3)$$
where $U(\omega)$ is the Fourier transform of $\mu$.
The expected size of the  error in  removing the $\ell'$th sinusoid 
from the estimate of the $\ell$th sinusoid,
$\widetilde{A_{\ell'}\cos(\phlp)}$, is given by (2.8).
The bias from the other line frequencies can be included as
a correction to (2.6). 

The kernels of (2.9) and Appendix B are designed to minimize
the total error under the assumption that the signal is well resolved.
These ``minimal loss'' kernels tend to have larger frequency sidelobes
because their design criterion does not explicitly penalize sidelobes.
Many other digital filters and differentiators [5] have been designed 
to have power spectra which decay rapidly away from zero frequency,
and thereby reduce the  interference terms because $U(\omlp -\oml ) \ll 1$.

If interference from sidelobes is significant, we replace the 
minimal loss kernels with kernel which satisfy  $U(\omlp -\oml ) \ll 1$ for
broad banded bias protection. Our particular choice is to construct
a kernel from the first $(p+1)$ sinusoidal tapers of [19].
The sinusoidal tapers are defined by
$v_n^{(k)}=\sqrt{{2 \over N+1}}\sin({\pi kn \over N+1})$
where $k$ is the taper number, $N$ is the length of the kernel
and $n=\ 1 \ldots N$. Imposing (2.3) and the condition that
the kernel vanish at the ends of its support gives $p+1$ conditions
and $p+1$ free parameters.

\  \\

\cl{VI. DISCUSSION}
\vspace{.05in}

In this article, we have treated time-frequency distributions
as an estimation problem for slowly varying sinusoids.
This model generalizes the Rife and Boorstyn problem [20] of estimating
a pure sinusoid in noise. This approach is valid and appropriate when
we know $a$ $priori$ that the signal consists of one or more coherent
signals in a noise background. When the signal is incoherent with
a slowly varying spectral density, the evolutionary spectrum of
Priestley [16] is the appropriate model. In [17], 
we use a two-dimensional cross-product kernel smoother to estimate the 
evolutionary spectrum, $S(\omega,t)$.

To our knowledge, none
of the previous instantaneous frequency estimators [2] 
include the effects of bias error 
{\it from the time variation of the frequency} in their analyses. The multiple
stage kernel estimators of the appendix yield the  optimal rate
of convergence for nonparametric estimation.
When the instantaneous frequency is known $a$ $priori$ to have a particular
parametric form such as a ``chirp'', more accurate estimators are possible.
The well-known Cramer-Rao bound of Rife and Boorstyn [20] applies when
the instantaneous frequency is time independent. 

We determine the instantaneous frequency
by estimating $e^{i\phtl_{}(t)}$ and $\partial_t e^{i\phtl_{}(t)}$ 
with kernel smoothers. The bias error is proportional
to $h^2\partial_t^3 e^{i\phtl_{}(t)}$. 
We demodulate by $e^{-i(\omega_o t+\phi_{o})}$
to reduce this bias. The variance of the
estimate of $\partial_t e^{i\phtl_{}(t)}$ scales as
${\sigma^2 \over |A|^2 N h^{3}}$. 
Minimizing the expected error  yields
the optimal kernel halfwidth, $h_{1,3} \sim  \left[{ \sigma^2 \over 
A^2N| \partial_t^3 (e^{i\phtl(t)} )|^2 } \right]^{1\over 7}$ and
$L^2(\widehat{\partial_t \phi}(t_j) ) \ \sim \
 |\partial_t^3 e^{i\phtl} |^{6 \over 7}
\left({ \sigma^2  \over A^2 \ N } \right)^{4 \over 7}$.

Our optimal kernel smoother
approach has two disadvantages. First, it is computationally more 
intensive than many of the alternative methods. Second,
the asymptotic expressions are based on a Taylor series expansion
of $e^{i\phtl_{}(t_j)}$ about $e^{i\phtl_{}(t)}$. If $h_t$ is the 
radius of validity of the third order expansion,
$$e^{i\phtl_{}(t_j)} \sim e^{i\phtl_{}(t)}+ {(t_j -t)}
\partial_t e^{i\phtl_{}(t)} + {(t_j -t)^2\over 2}
\partial_t e^{i\phtl_{}(t)}
+ {(t_j -t)^3\over 6}\partial_t e^{i\phtl_{}(t)} + o({(t_j -t)^{3+\delta}})
\ \ , \ $$ 
our analysis shows that the optimal halfwidth is given by (3.4) if $h_o$
is less than $h_t$. Since the optimal halfwidth scales as the $1/7$ root
of the signal to noise ratio divided by the number samples per characteristic
time, we are often in the limit where $h_o > h_t$. In this case, our analysis
shows only that the best kernel halfwidth is greater than or equal to $h_t$.
The failure of the Taylor series approximation often corresponds 
to an order one phase difference between
$e^{i\phtl_{}(t_j)}$ and $e^{i\phtl_{}(t)}$. Thus
the lower bound on the kernel halfwidth is useful and is often close to
optimal value.

 \ \\

\noindent
{APPENDIX A: DATA ADAPTIVE MULTIPLE STAGE KERNEL ESTIMATORS}
\vspace{.1in}

In this appendix,
we construct multiple step kernel estimators which have optimal
relative convergence rates.   
We return to the case of a nonmodulated real signal ($\phi(t) \equiv 0$). 
We consider data adaptive estimators which
estimate $g^{(q)}(t)$ in the final stage
with a kernel of order $(q,p)$ where the kernel parameters are determined
with a kernel pre-estimate of $g^{(p)}(t)$ of order $(p,p+2)$. 
The more accurate the estimate of $g^{(p)}(t)$ is, the closer the
expected loss of the ``plug-in'' kernel estimator will be to
the optimal value with known $g^{(p)}(t)$. 
If the estimated value of $\widehat{g^{(p)}}(t)$
differs from $g^{(p)}(t)$ by $O(N^{-\alpha})$, then 
$${\rm E}\left[|\hat{g}(t|\hh_{0,q}) - g(t)|^2 \right] \  
\siml \ \left( 1+ O(C_r^2 N^{-2\alpha})\right)
{\rm E}\left[|\hat{g}(t|h_{0,q}) - g(t)|^2\right]
\ ,\eqno(A1)$$
where $h_{0,q}$ is given by (2.7) and  $\hh_{0,q}$ is 
its empirical estimate. We say that $C_r N^{-\alpha}$ is
the relative convergence rate of the kernel halfwidth estimate.
(The convergence is relative to the rate with the known, optimal value of
$h$.)
If the relative convergence rate tends to zero as $N\rightarrow \infty$,
then the estimate is asymptotically efficient. 

To achieve the optimal rate of
convergence, the ``pre-estimates'' of $\partial_t^p g(t)$
are estimated with a different kernel length than the estimate of $g(t)$. 
Equation (2.7) shows that 
the optimal $h$ scales as $N^{-1/5}$ for kernels
of order $(0,2)$ and $(2,2)$, and that $h$ scales as 
$N^{-1/9}$ for kernels of order $(0,4)$ and $(2,4)$. 
For pre-estimates with kernels of order $(p,p+2)$, the relative convergence
rate is $O(N^{-2\over(2p + 5)})$. 

In the first step of any multistep estimation scheme, the kernel halfwidth
for the next step needs to be selected. There are three common
methods to initialize the kernel smooother:
characteristic time scale initialization, parametric fit initialization, and
goodness of fit initialization.
In the characteristic time scale initialization, the signal is assumed to
have a characteristic amplitude, $\Abr$, and to vary on a  
characteristic time scale, $\tau$, 
where $\Abr$ and $\tau$ are given $a$ $priori$. 
In the initial halfwidth estimate, $\Abr / \tau^p$ is sustituted for 
$g^{(p)}(t)$ in (2.7).

In the parametric fit initialization, $g(t)$ is fit with a prescribed 
functional form with a small number of free parameters.
The parametric fit is then substituted into (2.7) to initialize the
kernel estimate.

In the goodness of fit initialization, $h$ is determined by
minimizing an  expression which includes the residual sum of squared
errors but which corrects for the number of degrees of freedom
which are used in a kernel smoother [9,14,16,18]. 

In general, both the 
characteristic scale initialization and the parametric fit initialization
produce order one errors in $\widehat{g^{(p)}} (t)$.  
The goodness of fit criteria 
have a slow relative rate of convergence, $O(N^{-1/10})$ 
[9]. 
In contrast, the pilot kernel estimator of order $(p,p+2)$ has a 
relative convergence rate of $O(N^{-2/9})$.
Therefore
we recommend a multiple stage  kernel estimator where $g^{(p)}$ is estimated
prior to the estimation of $g(t)$.

For simplicity, we consider        
a two stage estimate with a characteristic
scale initialization. 
We begin the adaptive estimate by selecting a global halfwidth, $h_{2,4}$,
using a characteristic scale initialization. 
The resulting estimate of $\partial_t^2 g$ achieves the optimal
convergence rate of $L^2(\widehat{\partial_t^2 g})\ \siml\ C_2N^{-4/9}$.
Because we use an arbitrary
ansatz, $\partial_t^4 g \sim \tau^{-4} \Abr$, in the optimal halfwidth
formula,  
the convergence rate differs from the optimal value by an order one factor.  

To estimate $\part_t g(t)$ (as in instantaneous frequency estimation)
we  begin with the ansatz that $\partial_t^5 g \sim  \Abr/\tau^{5}$,
We insert this  ansatz into (2.7) to determine a halfwidth to estimate
$\part_t^3 g(t)$ using a kernel of order (3,5). 
We then use the  estimate of $\part_t^3 g(t)$ to 
determine a halfwidth for a kernel of order (1,3). 

Our estimate of $g(t)$ achieves the optimal
convergence rate of $L^2(\widehat{g}(t))
\siml CN^{-4/5}$ 
and the optimal relative convergence rate of $N^{-4/9}$.
A similar, slightly more elaborate, adaptive estimator
was proposed by M\"uller \& Stadtm\"uller (M-S) [15]. M-S begin by determining 
a global halfwidth for the (2,4) kernel using the Rice criterion.
Goodness of fit initializations improve on the characteristic time scale
initialization
by selecting an asymptotically efficient  global halfwidth
for the estimate of $\partial_t^2 g{(t)}$.
The M-S scheme is actually a three stage estimator and the computational
effort required on the initial step can be large.

Since $g(t)$ is $C^p$, we want $\hat{g}$ to be smooth as well. 
The kernel halfwidth of (2.7) using the ``plug-in'' derivative estimate
in an infinite kernel halfwidth. Also, $\widehat{\part_t^p g}$ is at best 
is continuous, but need not be smooth. 
Therefore we convolve
$|\widehat{\partial_t^p g}|^2$ with a regular kernel, $G(\cdot)$, with
$\int G(s)ds =1$. We choose the halfwidth of $G$ to be 
$\hb \equiv max_t\{ \hh(t)\}$. To apply our estimate of $g^{(p)}(t)$ to
(2.7), we make the substitution,
$$|\widehat{\partial_t^p g}|^2 \rightarrow  
{1\over 2{\hb}}\int_{-\hb}^{\hb} G({s \over \hb}) 
|\widehat{\partial_t^p g}(t-s)|^2 ds
\ .\eqno(A2)$$
This smoothed estimate is asymptotically equal to $|\partial_t^p g(t)|^2$,
 but robustifies the empirical halfwidth
for moderate values of $Nh$. When
$\partial_t^p g(t)$ nearly vanishes, 
the smoothing in (A2) models the effect of the higher order bias terms.
In [15], the robust estimator of 
$\partial_t^2 g$ 
in (A2) is replaced by a simpler, but less accurate upper cutoff.

\ \\

\cl{APPENDIX B:  OPTIMAL KERNEL SHAPES}


In (2.6), the expected loss is a quadratic function of the
kernel, $\mu$. For a fixed kernel width, 
we can minimize the expected
loss subject to the constraints that $\mu$ is of type $(q,p)$.
For a given, positive definite, symmetric matrix, $\Rbr$,
we define the
minimal $\Rbr$ kernel of order $(q,p)$, $\muRb$ as the minimizer of
$$
\muRb^T\Rbr \muRb + \sum_{m=0}^{p-1} \alpha_{m}
(\muRb\cdot\sbf_{m} - \delta_{m,q} )
,\eqno (B1)
$$
where $\alpha_{m}, \ m= \ 0 \ldots p-1$ are $p$ Lagrange multipliers
and $T$ denotes ``transpose''.
When $\Rbr \equiv \sigma^2\Ibf$, we call $\muR$ the
minimal variance kernel of type $(q,p)$ and the solution is given in
[13]. 
The approximate expected loss is given by:
$$ L(\muRb)\  \siml\  
\muRb^T\Rbf \muRb + |\partial_t^p g(t)|^2
|\muRb\cdot\sbf_{p}|^2
\ .\eqno (B2)$$
If both $\Rbf$ and $\partial_t^p g(t)$ 
are given, the minimal loss kernel, $\mubf_L$,
corresponds to the choice of 
$\Rbr = \Rbf \equiv \Rbf + |\partial_t^p g(t)|^2
\sbf^{(p)}\sbf^{(p)^T}$. Thus the expected loss functional differs from the
minimum variance functional by a rank one perturbation.

Thus the minimal $\Rbr$ norm kernel satisfies
$$
\Rbr \muRb \  =\ - \sum_{m=1}^{p-1} \alpha_{m}\sbf^{(m)}
,\eqno (B3)
$$
with the linear constraints of (2.3).
We define
the $N \times p$ matrix,
$\Sbf{(p)} \equiv (\sbf^{(0)},..,\sbf^{(p-1)})$; i.e.
$\Sbf{(p)}$, the matrix of the first $p$ moment vectors,
$\sbf^{(m)}, \  0\le m < p $.   
We  also define the $p$ vector, $\ebf_{q}^{(p)}$, to be
the unit $p$ vector in the $q$ direction, and
$\albf$ to be the $p$ vector of Lagrange multipliers.
The solution of (B2) is 
$$
\Sbf^T \Rbr^{-1} \Sbf \albf \ =\ - \ebf_{q}^{(p)}
,\eqno (B4)
$$
$$
 \muRb \ =\  \Rbr^{-1} \Sbf ( \Sbf^T \Rbr^{-1} \Sbf )^{-1}\ebf_{q}^{(p)}
\ .\eqno (B5)
$$
Thus the minimal loss is 
$\ebf_{q}^{(p)^T} ( \Sbf^T \Rbr^{-1} \Sbf )^{-1}\ebf_{q}^{(p)}$
In [21], the optimal kernel shapes given by (B5) are explicitly evaluated
in the large sample limit.

\ns

\cl{ACKNOWLEDGEMENTS}

We thank L. Cohen, C. Hurvich and  A. Sidorenko for useful discussions. 
The valuable comments of the referees are also appreciated.
This work was funded by the U.S. Department of Energy.


\cl{REFERENCES}
\begin{enumerate}
\item{N.~S.~Altman, 
``{Kernel smoothing with correlated errors},'' 
{\it J.~Amer.~Stat.~Assoc.} {vol.~85}, pp.~749-759, Sept.~1990.}


\item{B.~Boashash, 
``Estimating and interpreting the instantaneous phase,
Pts.~I \& II,''
{\it  Proc.~I.E.E.E.} {vol.~80}, pp.~520-570,  Apr.~1992.}

\item{L.~Cohen, 
``Time-frequency distributions- a review,''
{\it  Proc.~I.E.E.E.} {vol.~77}, pp.~941-981, July 1989.}

\item{L.~Cohen and C.~Lee, 
``Local bandwidth and optimal windows for the short time Fourier transform,''
in {\it Advanced algorithms and architectures for signal processing IV}, 
{\it  Proc.~S.P.I.E.} {vol.~1152}, pp.~401-425,  1990.}

\item{S.~C.~Dutta Roy and B.~Kumar,
``Digital Differentiators,''
{\it Handbook of Statistics}, vol.~10, pp.~127-158,
{N.~K.~Bose and C.~R.~Rao eds.},
 New York: {North Holland Pub.}, 1993. }

\item{H.~Ge and D.~W.~Tufts,
``Estimating the frequencies of two sinusoids using only the phase angles
of complex-valued data,''
Submitted for publication.}

\item {U.~Grenander and M.~Rosenblatt, 
{\it Statistical analysis of stationary time series.} 
New York: {Wiley}, 1957. }

\item{W.~Hardle,  {\it Applied nonparametric regression.}
Cambridge: {Cambridge University Press}, 1990. }

\item{ W.~Hardle, P.~Hall and S.~Marron,  
 {``How far are automatically chosen smoothing 
parameters from their optimum?''}
{\it J.~Amer.~Stat.~Assoc.} {vol.~83}, pp.~86-95, March 1988.}

\item{F.~Hlawatsch \& G.~F.~Boudreaux-Bartels,  
``Linear and quadratic time-frequency signal representations,''
{\it  I.E.E.E.~Signal Processing Mag.} {vol.~9}, 21-67, Apr.~1992.}

\item{S.~Kay,
``Estimating the frequencies of a noisy sinusoid by linear regression,''
{\it I.E.E.E.~Trans.~in Acoust.~Speech, Signal Processing} 
{vol.~37}, pp.~1987-1990, Dec.~1989.}

\item{B.~C.~Lovell and R.~C.~Williamson,
``The statistical performance of some instantaneous frequency estimators,''
{\it I.E.E.E.~Trans.~in Signal Processing} 
{vol.~41}, pp.~1708-1723,   July 1992.}

\item{H.~G.~M\"uller,
 ``Weighted local regression and kernel methods 
for nonparametric curve fitting,''
{\it J.A.S.A.} {vol.~82}, pp.~231-238, March 1987. }

\item{H.~G.~M\"uller, 
{\it Nonparametric regression analysis of longitudinal data.}
Berlin: Springer Verlag, 1980.}

\item{H.~G.~M\"uller and U.~Stadtm\"uller, 
 {``Variable bandwidth kernel estimators of regression curves,''}
{\it Annals of Statistics} {vol.~15}, pp.~182-201, Jan.~1987.}

\item{M.~B.~Priestley,  {\it Spectral analysis and timeseries.} 
Ch.~11,  New York: { Academic Press},  1981.}

\item{K.~S.~Riedel, 
``Data-based kernel estimation of evolutionary spectra,''
{\it I.E.E.E. Trans.~in Signal Processing} 
{vol.~41}, pp.~2439-2447, July, 1993. 
Also {\it Proc.~I.E.E.E.-S.P.~Int.~Symp.~on Time-Freq.~\& Time-Scale},}
Victoria, British Columbia, pp.~273-276,  1992.

\item{ K.~S.~Riedel and A.~Sidorenko, ``Data Adaptive Kernel Smoothers,''
To be published in {\it Computers in Physics},  1994.}

\item{ K.~S.~Riedel and A.~Sidorenko, (1994). 
``Minimum bias multiple taper spectral estimation,''
Submitted for publication.}

\item{D.~C.~Rife and R.~R.~Boorstyn, 
``Single-tone parameter estimation from discrete-time observations,''
{\it  I.E.E.E.~Trans.~on Information Th.} {vol.~20}, pp.~591-598,
1974.}  

\item{A.~Sidorenko and K.~S.~Riedel, 
``Optimal boundary kernels and weightings for local polynomial regression,''
Submitted for publication, 1994.} 

\item
{ C.~J.~Stone, 
{``Optimal global rates of convergence for nonparametric regression,''}
{\it  Annals of Stat.} {vol.~10}, pp.~1040-1053, 1982.}

\item{D.~J.~Thomson  and A.~D.Chave, 
{``Jackknife error estimates for spectra, 
coherences and transfer functions,''} 
in {\it Advances in spectrum analysis},
(S.~Haykin ed.) Ch.~2, pp.~58-113, 
Prentice-Hall, New York, 1990.}

\item{D.~W.~Tufts and R.~Kumarsen,
``Estimation of frequencies of multiple sinusoids: making linear prediction
perform like maximum likelihood,'' 
{\it  Proc.~I.E.E.E.} {vol.~70}, pp.~975-989,  Sept.~1982.}

\end{enumerate}

\end{document}